\documentclass[
showpacs,
floatfix,
aps,
prl,
twocolumn,
superscriptaddress,
amssymb,
]{revtex4-1}

\usepackage{amssymb}
\usepackage{latexsym}
\usepackage[normalem]{ulem}
\usepackage{graphicx}
\usepackage{amsmath}
\usepackage{dcolumn}
\usepackage{amsfonts}
\usepackage{bm}
\usepackage{color}

\usepackage{siunitx}
\usepackage{hyperref}
\usepackage{physics}

\newcommand{\be}{\begin{equation}}
\newcommand{\ee}{\end{equation}}
\newcommand{\bea}{\begin{eqnarray}}
\newcommand{\eea}{\end{eqnarray}}

\def\gm{$\Gamma$ }
\def\wse2{WSe$_2$}
\def\c12{$c_{12}$}

\begin{document}

\title{Contrasting \gm and $K$ valley moiré physics in twisted monolayer/bilayer \wse2}

\author{Jackson Kuklin}
\affiliation{Department of Physics and Astronomy, University of California, Los Angeles, CA 90095, USA}
\author{Ning Mao}
\affiliation{Max Planck Institute for Chemical Physics of Solids, 01187, Dresden, Germany}
\author{Milan Sharma Mandigo-Stoba}
\affiliation{Department of Physics and Astronomy, University of California, Los Angeles, CA 90095, USA}
\author{Edgar Elias}
\affiliation{Department of Physics and Astronomy, University of California, Los Angeles, CA 90095, USA}
\author{Tianci Song}
\affiliation{Department of Physics and Astronomy, University of California, Los Angeles, CA 90095, USA}
\author{Connor Engel}
\affiliation{Department of Physics and Astronomy, University of California, Los Angeles, CA 90095, USA}
\author{Pola Pietrzkowski}
\affiliation{Department of Physics and Astronomy, University of California, Los Angeles, CA 90095, USA}
\author{Robert Boyd}
\affiliation{Department of Materials Science and Engineering, University of Wisconsin-Madison, Madison, Wisconsin 53706, USA}
\author{Kenji Watanabe}
\affiliation{Research Center for Functional Materials,
National Institute for Materials Science, 1-1 Namiki, Tsukuba 305-0044, Japan}
\author{Takashi Taniguchi}
\affiliation{International Center for Materials Nanoarchitectonics,
National Institute for Materials Science,  1-1 Namiki, Tsukuba 305-0044, Japan}
\author{Daniel Rhodes}
\affiliation{Department of Materials Science and Engineering, University of Wisconsin-Madison, Madison, Wisconsin 53706, USA}
\affiliation{Department of Physics, University of Wisconsin-Madison, Madison, Wisconsin 53706, USA}
\author{Yang Zhang}
\affiliation{Department of Physics and Astronomy, University of Tennessee, Knoxville, Tennessee 37996, USA}
\affiliation{Min H. Kao Department of Electrical Engineering and Computer Science, University of Tennessee, Knoxville, Tennessee 37996, USA}
\author{Qianhui Shi}
\affiliation{Department of Physics and Astronomy, University of California, Los Angeles, CA 90095, USA}

\date{\today}

\begin{abstract}

Electronic orbital character plays a central role in determining electronic correlations, spin--orbit coupling, dimensionality, and ultimately the quantum phases of condensed-matter systems. 
Two-dimensional moiré materials have emerged as highly tunable platforms for exploring correlated phenomena, but the role of orbital degrees of freedom remains largely unexplored.
Here, we identify twisted monolayer/bilayer WSe$_2$ as a platform in which displacement-field tuning enables moiré physics to be realized in both the $K$ and \gm valleys. 
The distinct orbital characters of these valleys give rise to contrasting correlated phases at moiré filling factors $\nu=-1$ and $\nu=-1/3$. 
At $\nu=-1$, the $K$-valley state is a weak insulator, consistent with an antiferromagnetic state near a van Hove singularity in the intermediate-coupling regime, similar to that observed in twisted bilayer WSe$_2$. In contrast, the \gm-valley state exhibits a pronounced Pomeranchuk effect, consistent with proximity to a Mott transition. At $\nu=-1/3$, the $K$ valley hosts a robust generalized Wigner crystal, whereas the \gm-valley state lies near the crystallization boundary and again exhibits a Pomeranchuk effect, with localization enhanced by increasing temperature or magnetic field.
Our work highlights the importance of orbital character in defining quantum phases in moiré systems, and identifies the \gm valley as a promising platform for exploring correlated phenomena near quantum phase transitions, where competing phases and enhanced fluctuations may give rise to unconventional phases.

\end{abstract}

\maketitle

The electronic orbital degree of freedom plays an essential role in shaping quantum phases in strongly correlated systems.
Prominent examples range from Mott insulators in transition metal oxides \cite{tokura2000} to ongoing efforts to understand the similarities and differences between nickelate and cuprate superconductors \cite{botana2020a,wang2024a}.
Two-dimensional moiré systems have emerged as a highly tunable platform for exploring correlated quantum phenomena, enabling the observation of states ranging from fractional anomalous Hall states to superconductivity \cite{balents2020,mak2022,cao}.
A defining strength of the 2D moiré system is its versatile tunability - the stacking order, twist angle, carrier density and displacement field allows control over topology, bandwidths and interaction strengths \cite{kennes2021}.
However, the orbital degree of freedom has received relatively little attention in moiré systems.

\begin{figure*}
    \centering
    \includegraphics{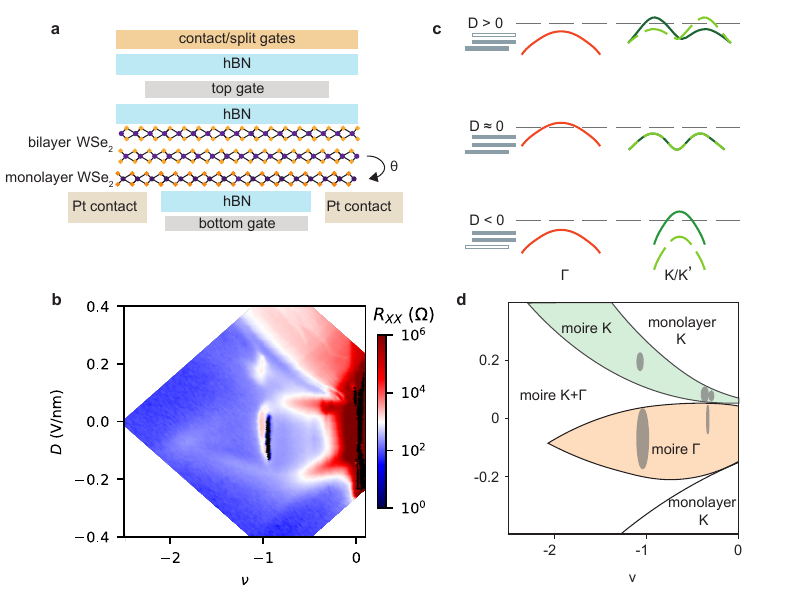}
    \caption{Overview of the device and valley population.
    (a) Schematic illustration of the device structure with a moiré interface between the lower two \wse2 layers.  
    (b) $R_{xx}$ vs $\nu$ and $D$ at $T =\SI{1.5}{K}$ and $B =\SI{0}{T}$.
    (c) Illustration of the relative alignment of the bands originating from the $K$ and $\Gamma$ valley at varying displacement fields. Solid (open) bars represent layers that are populated (unpopulated). 
    Dashed lines represent the Fermi levels at low densities. 
    (d) Schematic phase diagram indicating the valley population in each regime.
    Prominent resistive states are highlighted in grey. 
    }
    \label{fig:Fig1}
\end{figure*}

Two dimensional transition metal dichalcogenides (TMDs) provide a unique platform to engage the tuning knob on the orbital character \cite{lei_moire_2025}.
In conventional correlated materials, the orbital degree of freedom is typically modified through chemical composition, high pressure or strain, which may introduce additional structural or disorder effects. The tunability usually involves changing the weight of certain orbitals rather than switching between bands of different orbital characters.
In contrast, in 2D TMDs, the orbital character is cleanly tied to the valley degree of freedom:
near the valence-band edge, the K/K$^\prime$ valley bands are primarily composed of the transition metal $d_{x^2-y^2}\pm i d_{xy}$ orbits, whereas the \gm valley bands contain substantial $d_{z^2}$ and chalcogen $p_z$ orbits \cite{fang_ab_2015}.
As a result, they bear different correlation, spin-orbital coupling and layer hybridization:
first, the effective mass in the \gm valley is larger than that in the $K$ valley \cite{movva2018b}, leading to flatter bands and potentially stronger correlation effects in moiré superlattice systems;
second, the spin–orbit coupling is much weaker in the \gm valley---the spins realize Heisenberg rather than Ising spins as in the $K$ valley;
third, due to the out-of-plane character, interlayer hybridization is much stronger in the \gm valley, whereas $K$ valley wavefunctions are more layer-localized. 
These different layer-hybridization properties provide a natural mechanism to tune the relative energy of these valleys and their population.
While in monolayers and most bilayers, the valence band maximum is at the $K$ valley, more interlayer hybridization in thicker layers raises the energy of the \gm valley to the band edge \cite{vos_atomic_1999, movva2018b}. 
In addition, the two valleys respond differently to an applied displacement field: the layer-localized $K$ valley bands shift strongly with the displacement field, while the \gm valley bands are much less affected \cite{movva2018b,olin2024a,wei2025}. 
As a result, a displacement field provides convenient tuning for the population of the two valleys.

To date, the study of the moiré physics in TMDs have mainly focused on the $K$ valley in twisted homobilayers or heterobilayers \cite{wu2019,regan2020,tang2020,wang2020,xu2020,kangDoubleQuantumSpin2024,xia2025a,guo2025,xia2026,caiSignaturesFractionalQuantum2023,zengThermodynamicEvidenceFractional2023,park2023,xu2023a,han2026}, while theoretical studies have predicted a rich variety of correlated phases in \gm valley-derived moiré bands \cite{xian2021,zhang2021electronic,angeli2021,pan2023}.
Spectroscopic studies \cite{pei2022,gatti2023} have resolved \gm-valley-derived moiré bands, revealing their flat dispersion and strong reconstruction by the moiré potential, and established the single-particle basis for \gm-valley moiré physics.
Recent experiments have begun to probe electronic correlations in this regime, revealing correlated insulators, charge-transfer physics involving nearby \gm- and K-derived bands \cite{foutty2023,wei2025,campbell2024}, a honeycomb-lattice Mott transition \cite{ma2025}, and suppressed magnetism \cite{jiao2026}.
However, direct experimental evidence for how the distinctive orbital character of the active \gm band reshape the correlated phases remains largely unexplored.

Here, we report a study on a twisted monolayer/bilayer \wse2 system, which, assisted by a displacement field, offers the opportunity to study the moiré physics in both the $K$ and \gm valley in the same device. 
Unlike the twisted double bilayer systems that recently accesses the \gm valley with a honeycomb lattice \cite{ma2025,wei2025}, the lack of $C_{2y}$ symmetry in the twisted monolayer/bilayer structure leads to a triangular moiré potential (see Supplementary Material).
Within the same structural moiré supperlattice, the distinct orbital characters of the two valleys give rise to different effective moiré potentials and ultimately distinct correlated phases at both integer and fractional moiré fillings. 
At moiré filling factor $\nu = -1$ ($\nu = n/n_M$, where $n$ is the carrier density with $n < 0$ for hole carriers and $n_M >0$ is the moiré density), 
in the $K$ valley, we observe a weak insulator that develops at low temperature;
in the \gm valley, we observe a Pomeranchuk effect \cite{rozen2021,saito2021,zhang2022c,holleis2025} - although the state is more insulating in a large temperature range, the resistance drops sharply below about 7 K.
At $\nu = -1/3$, the $K$ valley hosts strong generalized Wigner crystals \cite{xu2020a,zong2025}, whereas the states in the \gm valley again shows Pomeranchuk effect at zero field, with crystalization enhanced by increasing temperature and a magnetic field. 
These results establish valley and orbital character as a powerful control parameter for correlated phases in moiré systems and opens new routes for distinct correlated states in the \gm valley.


\begin{figure*}
    \centering
    \includegraphics{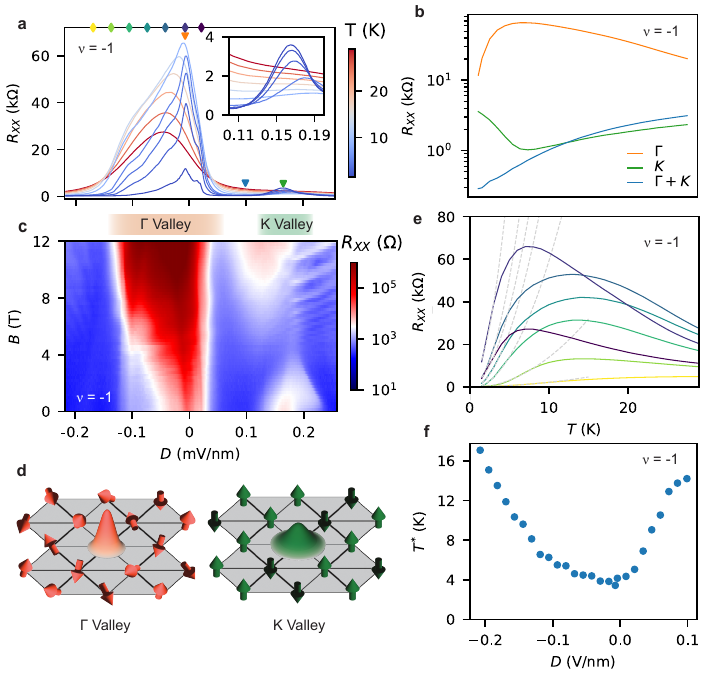}
    \caption{
    Contrasting correlated states in \gm and $K$ valley at moiré filling $\nu =-1$. 
    (a) $R_{xx}$ vs $D$ at $\nu = -1$ and different temperatures.  \gm and $K$ valley occupy different $D$ ranges, as marked. Inset is a zoom-in of the $K$ valley region. 
    (b) $R_{xx}$ vs $T$ at $D = -7$, $99$, $\SI{166}{mV/nm}$, as marked by the colored triangles in (a).
    (c) $R_{xx}$ vs $D$ and $B$ at $\nu = -1$. 
    (d) Schematic drawing of the Hubbard model on a triangular lattice; the wavefunctions are more localized in the \gm valley than in K, and the spin is more Heisenberg-like rather than Ising-like. 
    (e) $R_{xx}$ vs $T$ at different displacement fields in the \gm valley as marked by the colored diamonds in (a). Dashed lines are fits of $R_{xx} = R_0 +AT^2$. A characteristic temperature $T^*$ is defined as where the fit deviates from the data by 10\%.
    (f) $T^*$ vs $D$.
    }
    \label{fig:Fig 2}
\end{figure*}

Fig. \ref{fig:Fig1}(a) shows a schematic of the device studied.
An as-exfoliated 2H-stacked bilayer \wse2 is rotated and stacked on a monolayer, with the adjacent layers at a small angle mismatch around 3.4\textdegree, forming an ABB$^\prime$ structure with a moiré superlattice between the BB$^\prime$ layers. 
A displacement field, controlled by the top and bottom gate voltages, is applied across the trilayer structure. 
Fig. \ref{fig:Fig1}(b) shows the resistance $R_{xx}$ vs. the displacement field $D$ and $\nu$ at $B = 0$ and $T = 1.5$ K. 
A schematic band diagram depicting the effect of the displacement field on valley occupation is shown in Fig. \ref{fig:Fig1}(c).
At small carrier densities, only the moiré band emerging from the \gm valley is populated at $D = 0$.
A displacement field in either direction raises the $K$ valley in energy relative to the \gm valley, and the $K$ valley becomes predominantly populated.

The effect of the displacement field is asymmetric with respect to its direction, as expected from the structural asymmetry in the twisted monolayer/bilayer, and is most clearly reflected in the evolution of the resistive states at $\nu = -1$:
at small $D$, the \gm valley moiré bands host resistive state at $\nu = -1$;
as $D$ biases the carriers towards the twisted interface, the carriers occupy the moiré bands formed by the hybridization of the $K/K'$ valleys of the two layers, another resistive state is observed at $\nu = -1$, with a clear separation from the \gm valley state;
as $D$ biases the carriers away from the twisted interface, the carriers experience a weaker moiré potential, and $\nu = -1$ becomes featureless.
A schematic phase diagram of the valley population versus $D$ and $\nu$ is shown in Fig. \ref{fig:Fig1}(d), where correlated insulating states emerge at commensurate fillings when a well-defined moiré band of either valley is populated.
The identification of different regimes of the phase diagram is also supported by first-principle calculations and experimental data on quantum oscillations at high field (see Supplementary Material).

\begin{figure*}
    \centering
    \includegraphics{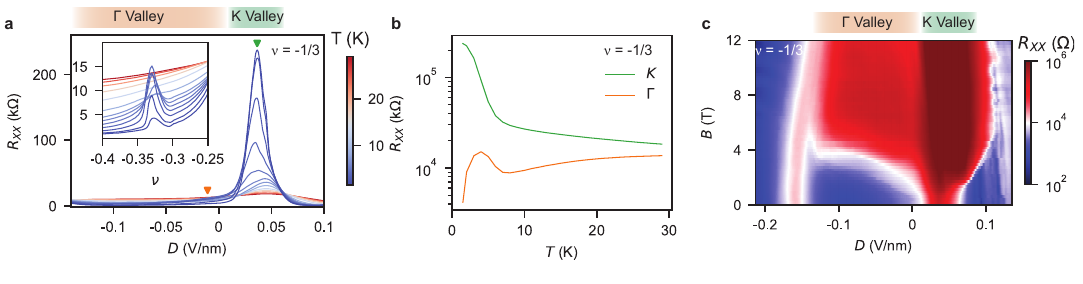}
    \caption{Contrasting correlated states in \gm and $K$ valley at moiré filling $\nu  = 1/3$.
    (a) $R_{xx}$ vs $D$ at $\nu = -1/3$ and different temperatures, as marked. Inset shows $R_{xx}$ vs $\nu$ at $D = \SI{-10}{mV/nm}$, highlighting maxima at $\nu=1/3$.
    (b) $R_{xx}$ vs $T$ at $D = -10$, $\SI{36}{mV/nm}$, as marked by the colored triangles in (a).
    (c) $R_{xx}$ vs $D$ and $B$ at $\nu = -1/3$. 
}
    \label{fig:Fig3}
\end{figure*}

The correlated states at $\nu = -1$ in the $K$ and \gm valley moiré bands show distinct properties.
Fig. \ref{fig:Fig 2}(a) shows the temperature dependence of the resistances at $\nu = -1$, with the $K$ and \gm valley states occupying different displacement field ranges, as marked.
In the $K$ valley regime, the resistance shows metallic behavior at high temperatures, and increases with decreasing temperature below about 7 K, indicating the development of an insulating state.
On the other hand, in the \gm valley regimes, the resistance is much higher and gradually increases with decreasing temperature in a large temperature range, but becomes less resistive upon further cooling below about 7 K, characteristic of a Pomeranchuk effect \cite{goldstein1959}. 
The distinct temperature dependencies at two characteristic displacement fields in these two regimes are shown by the orange and green curves in Fig. \ref{fig:Fig 2}(b).
At an intermediate displacement field in between these two points, where both $K$ and \gm valley are populated, the system exhibits metallic behavior in the whole temperature range, as shown by the blue curve in the same plot.

Fig. \ref{fig:Fig 2}(c) further shows the contrasting magnetic-field responses of the $K$- and \gm-valley states. 
In the $K$ valley, the insulating state is quickly suppressed around $B=4$ T; at higher fields, quantum oscillations emerge, closely resembling those in twisted bilayer WSe$_2$ \cite{han2025}.
In contrast, the \gm valley state become increasingly resistive at high magnetic fields. 

 While both of the $K$ and \gm valley realize a Hubbard model on a triangular lattice \cite{wu2018b, kumar2025} at $\nu = -1$, the different behaviors in the two valleys originate from the different energy scales and spin properties, as illustrated in Fig. \ref{fig:Fig 2}(d).
The \gm valley has strong out-of-plane orbital character, so interlayer hybridization is large and strongly stacking-dependent. This amplifies the moiré potential and, together with the heavier \gm effective mass, produces more localized moiré Wannier orbitals in the superlattice.
As a result, the smaller tunneling amplitude $t$ and larger onsite interaction energy $U$ leads to a larger $U/t$ in the \gm than in the $K$ valley. 
In the $K$ valley, the system lies in the intermediate regime, where a correlated insulating state only appears close to the van Hove singularity \cite{wang2020,guo2025}.
The $\nu = -1$ state in our device is similar to those in twisted bilayer \wse2, and the weak insulating state observed at zero field and its rapid suppression by magnetic field is consistent with the zero-field state being an inter-valley coherent antiferromagnetic state \cite{guo_angle_2026}, while the magnetic field splits the $K$ and $K'$ moiré bands and leads to exciton insulators, as suggested by studies in twisted bilayer \wse2 \cite{han2025} (see Supplementary Material for comparison with the $\nu = -1$ state in twisted bilayer \wse2). 
On the other hand, in the \gm valley, the $U/t$ is estimated to be around 8 (see Supplementary Material) at the twist angle in our device, and place it very close to the onset of Mott transition.

Near the Mott transition, theoretical works \cite{wietek2021a} have predicted a Pomeranchuk effect \cite{goldstein1959}: 
the insulating state, with free magnetic moments, has a higher entropy than the metallic state that is close in energy; elevated temperatures then enhance the entropy contribution in the free energy, and promote the localized states.
In addition, the weak spin-orbit coupling in the \gm valley realizes a Hubbard model with SU(2) symmetry, while in the $K$ valley, the spin interactions are more Ising-like due to spin-valley locking and complex, valley-contrasting moiré hoppings \cite{devakulMagicTwistedTransition2021a}.
The richer spin fluctuations in the nearly SU(2)-symmetric \gm valley may allow substantial spin entropy to persist over a broader temperature range, further promoting the Pomeranchuk effect.
The magnetic-field response in Fig.2(c) provides a complementary manifestation of the same competition: localized magnetic moments have a larger spin susceptibility than the nearby metallic state, so their energy are lowered more by the magnetic field and therefore increasingly favored.

Furthermore, even in the regime that $\dv*{R}{T}$ indicates metallic behavior, the temperature dependence deviates from a Fermi liquid behavior starting from only a few kelvin, as shown in Fig. \ref{fig:Fig 2}(e) and (f).
Possible origins for this behavior include competing phases at low temperatures, or the emergence of a substantial entropy contribution even at low temperatures.
The exchange energy scale, estimated from calculations for the \gm band, is roughly $J=4t_1^2/U \approx \SI{7}{K}$.
At temperatures well below this scale, the exchange correlation between the local mangetic moments becomes important and suppresses the entropy which will finally lead to an ordered magnetic state with zero entropy at zero temperature.
One notable exception is a quantum spin liquid, which can retain substantial entropy without developing long-range magnetic order.
Such phases have been widely predicted near the Mott transition in a triangular-lattice Hubbard model \cite{senthil_theory_2008,balents_spin_2010,mishmash_continuous_2015,szasz2020,drescher2023a,wietek2021a}.
In addition to the geometric frustration of the triangular lattice, a sizable next-nearest-neighbor hopping $t_2$ relative to the nearest-neighbor hopping $t_1$, $|t_2/t_1| \approx 0.15$  (see Supplementary Material)  introduces kinetic frustration, which could further destabilize magnetic order and promote spin-liquid phases \cite{tocchio2020a,zhu2015,hu2015}.
Future experiments to characterize the temperature dependence of the entropy may help to determine whether the entropy at $\nu = -1$ is from weakly-correlated magnetic moments or spin liquid phases. 

The resistance of the \gm-valley state also evolves substantially with displacement field. 
While the intrinsic \gm-valley moiré bandwidth changes only weakly with $D$, the energy separation to the nearby unoccupied K-derived moiré and non-moiré band varies more strongly, which likely contribute to the observed $D$-dependence.
The proximity of these bands may provide an additional route to tune the effective interaction parameters of the \gm-valley state with displacement field.
Nevertheless, this band separation remains larger than the relevant thermal energy scale over the temperature range of the Pomeranchuk behavior, such that a single-band \gm-valley Hubbard model remains a reasonable low-energy description (see Supplementary Materials).


Next, we turn to the correlated states at fractional moiré fillings \cite{regan2020,xu2020}, which also manifest distinct behavior in the two valleys.
The correlated states, appearing at fractional filling in a Chern-zero band, is due to the formation of charge density waves (often termed as generalized Wigner crystals, GWC) due to long range Coulomb interactions \cite{matty2022,kumar2025}.
In the $K$ valley regime, we observe strongly resistive states at $\nu = -1/3$ and $-1/4$; in the \gm valley we observe a weaker $\nu = -1/3$ state while no states are discernible at $-1/4$.
In Fig. \ref{fig:Fig3}(a) we plot the resistances along $\nu = -1/3$ at different temperatures with the $K$ and \gm valley states occupying different displacement field ranges, as marked.
While the $K$ valley states manifest a standard insulating behavior, the \gm valley $\nu = -1/3$ state is much less pronounced.
Its presence is more clearly seen in the density dependence at a fixed displacement field, shown in the inset of Fig. \ref{fig:Fig3}(a), where a distinguished peak appears at $\nu = -1/3$.
Interestingly, similar to the $\nu = -1$ state, the $\nu = -1/3$ state in the \gm valley exhibits a Pomeranchuk-like temperature dependence: the resistance peak develops below about 7 K, then decreases with decreasing temperature, showing a maximum at intermediate temperature.
The resistance vs temperature for two representative displacement fields in the $K$ and \gm valley are shown in Fig. \ref{fig:Fig3}(b).
All these observations suggests that the charge density waves are stabilized in the $K$ valley, whereas the \gm valley represents a regime that is on the verge of the transition out of the GWC.
The contrast between the GWC states in the $K$ and \gm valley is consistent with a weaker long-range Coulomb interaction and stronger Wannier orbital localization in the \gm valley.

A Pomeranchuk effect has been proposed \cite{spivak2004} for Wigner crystals closely competing with a liquid phase, with recent experimental indications in the dilute limit in the absence of a superlattice \cite{sung2025}.
It has a similar origin to the Pomeranchuk effect close to a Mott insulator: the large spin entropy of solid phase makes it more favorable at a higher temperature. 
A distinction between the states at $\nu = -1/3$ and $\nu = -1$, however, is the different temperature scales for the resistance variations.
At $\nu = -1$, the insulating behavior persists well above 30 K, while it gradually gives in to a metallic-like behavior below around 8K.
On the other hand, at $\nu = -1/3$, the insulating behavior only develops below 7 K, and quickly turns over to a sharp drop below 4 K. 
This contrast reflects the different energy scales for the Mott insulator and the GWC states: charge localization is primarily governed by the onsite interaction for the former and the long range Coloumb interaction for the latter, while the exchange interaction strengths---which decides when the entropy becomes important---is also smaller for the GWC states as the carriers are farther apart.

Under a perpendicular magnetic field, the \gm valley -1/3 state demonstrates a sharp turn-on of insulating state around 4 T.
Similar to the $\nu = -1$ state, this is again consistent with a magnetic field induced preference towards the solid phase due to its larger susceptibility. 
On the other hand, the $\nu = -1/3$ state in the $K$ valley persists up to high field, consistent with it being in the deep crystallization regime.

\begin{figure}
    \centering
    \includegraphics{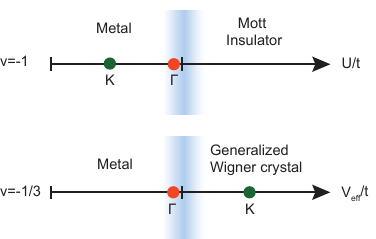}
    \caption{
    Schematic phase diagram of the metal-to-Mott insulator transition at $\nu = -1$ and the metal to GWC transition at $\nu = -1/3$. 
    The $K$ valley is in the intermediate interaction regime in terms of $U/t$ at $\nu = -1$ and in the robust GWC regime at $\nu = -1/3$, while the \gm valley states are close to the transition for both.
}
    \label{fig:Fig4}
\end{figure}

We summarize the contrasting behavior of the $K$ and \gm valley at $\nu = -1$ and $\nu = -1/3$ in Fig. \ref{fig:Fig4}.
As the Mott transition at $\nu = -1$ and the transition to the GWC at $\nu = -1/3$ are controlled by different energy scales, the two valleys exhibit opposite hierarchy.
Compared to the $K$ valley, the \gm valley lies closer to a Mott insulator, but is less favorable for GWC formation.
Remarkably, in our 3.4 degree twisted monolayer/bilayer \wse2, the \gm valley states at both fillings lie near their respective phase boundaries.
This provides a unique platform for exploring quantum phases in regimes of strong competition and enhanced fluctuations near phase boundaries.




\section{Acknowledgements}
Y.Z. was supported by the Max Planck partner lab for quantum materials. N.M. acknowledge financial support by the Deutsche Forschungsgemeinschaft (DFG, German Research Foundation) through the Würzburg-Dresden Cluster of Excellence ctd.qmat – Complexity, Topology and Dynamics in Quantum Matter (EXC 2147, project-id 390858490). DR acknowledges the Office of Naval Research (N000142512276). RB acknowledges the Army Research Laboratory under Cooperative Agreement Number W911NF-24-2-0054. The views and conclusions contained in this document are those of the authors and should not be interpreted as representing the official policies, either expressed or implied, of the Army Research Laboratory or the U.S. Government. The U.S. Government is authorized to reproduce and distribute reprints for Government purposes notwithstanding any copyright notation herein.


\end{document}